\documentstyle[preprint,aps,epsfig]{revtex}
\begin{document}
\title{Temperature and Angular Dependence of the
Magnetoresistance in Low Dimensional Organic Metals}
\author{J. S. Qualls$^1$, J. S. Brooks$^2$, S.
Uji $^3$, T. Terashima$^3$, C. Terakura$^3$, H. Aoki$^3$, L. K. Montgomery $^4$}
\address{$^1$Wake Forest University, Winston-Salem, NC 27109, USA}
\address{$^2$National High Magnetic Field Laboratory, Florida State University, Tallahassee, FL 32306, USA}
\address{$^3$National Research Institute for Metals, Tsukuba, Ibaraki 305, Japan}
\address{$^4$Department of Chemistry, Indiana University, Bloomington, IN 47405, USA}
\maketitle
\date{Received: \today }

\begin{abstract}Detailed studies of the magnetoresistance of
$\alpha$-(ET)$_{2}$KHg(SCN)$_{4}$ and $\alpha$-(ET)$_{2}$TlHg(SCN)$_{4}$
as a function of temperature, magnetic field strength, and field
orientation are reported. Below 15 K, the temperature dependence of the magnetoresistance
is metallic (dR/dT $>$ 0) for  magnetic field orientation corresponding
to an angular dependent magnetoresistance oscillation (AMRO) minimum and 
nonmetallic (dR/dT $<$ 0) at all other field orientations. 
We find that this behavior can be explained in terms of semiclassical models without the 
use of a non-Fermi liquid description. The alternating temperature dependence (metallic/nonmetallic)
with respect to field orientation
is common to any system with either quasi-one or two-dimensional AMRO.
Furthermore, we report a new metallic property of the
high field and low temperature regime of $\alpha$-(ET)$_{2}$MHg(SCN)$_{4}$ (where M = K, Rb, or Tl)
compounds.\end{abstract}


%\twocolumn

{\bf I. INTRODUCTION}

Recently, non-Fermi liquid behavior has been proposed to describe transport
properties of the quasi-one-dimensional organic conductor
(TMTSF)$_{2}$PF$_{6}$.\cite{kriza,strong,zheleznyak,lebed,chashech} This assignment
is based on the temperature and angular dependence of the magnetoresistance in 
tilted magnetic fields. A characteristic signature
of the angular dependent magnetoresistance of (TMTSF)$_{2}$PF$_{6}$ is its oscillatory
nature, where at ``magic angles'' the resistance exhibits sharp dips against
a broad $cos(\theta )^{\gamma}$ background.\cite{osada,kang} This feature is thought to arise from the
warped open orbit Fermi surface sheets. As the field
strength is increased along the second most conducting ${\bf b}$ axis, the effective electronic dimensionality is decreased.
For temperatures below $\approx$ 20 K this reduction in dimensionality leads to a decoupling of the layers (or a loss of phase coherence for electronic states between planes),
i.e. non-Fermi liquid behavior and a nonmetallic (dR/dT $<$ 0)
temperature dependence.\cite{zheleznyak,chashech}  Below 8 K, for fields orientations generating dips in the magnetoresistance,
a metallic (dR/dT $>$ 0) temperature dependence is recovered.
Away from the magic angles the temperature dependence changes slightly,
however it remains non-metallic (dR/dT $<$ 0). This metallic/nonmetallic behavior is explained in terms of a restoration of the interplane coupling
at the magic angles and a second decoupling at all other orientations due to fields along the ${\bf c}$
axis.\cite{zheleznyak,chashech}

Similar to (TMTSF)$_{2}$PF$_{6}$, the material $\alpha$-(ET)$_{2}$KHg(SCN)$_{4}$
displays angular dependent magnetoresistance oscillations (AMRO) and metallic/non-metallic temperature dependence of the magnetoresistance.
Does this indicate that $\alpha$-(ET)$_{2}$KHg(SCN)$_{4}$ experiences non-Fermi liquid transport?
If not, what is the relationship between the temperature dependence and AMRO?

At room temperature, $\alpha$-(ET)$_{2}$MHg(SCN)$_{4}$, where M = K, Tl, 
or Rb has a Fermi surface defined
by the coexistence of a quasi-two-dimensional cylinder and quasi-one-dimensional
sheets.\cite{mori,ducasse} By decreasing the temperature,
a Fermi surface reconstruction occurs at $T_{DW}$ (where $T_{DW}$ = 8 K, 10 K, or 12 K,
for M = K, Tl, or Rb respectively) and a new electronic ground state
develops.\cite{ishiguro} The true nature of the low temperature ground state is a subject of contemporary debate, however 
approaches assuming a charge density wave framework appear very promising.\cite{biskup,sasaki,spuds,qualls}
Figure 1 is a schematic $B-T$ phase diagram for
the $\alpha$-(ET)$_{2}$KHg(SCN)$_{4}$ compound.\cite{biskup,kartsovnik2} For the field perpendicular to the most conducting
plane, the electronic structure can be separated into three regimes;
the normal state, density wave one (DWI), and density wave two
(DWII). Evidence of a third phase has been observed for field orientations
near the most conducting plane.\cite{qualls} The electronic structure of this new phase
appears to be very similar to that of DWI. To simplify the following
arguments it will not be examined in this work. Each electronic regime of $\alpha$-(ET)$_{2}$KHg(SCN)$_{4}$ has its
own characteristic dependence of the magnetoresistance with respect to
field orientation, which is indicated
in Fig. 1 as either quasi-one or quasi-two-dimensional. At low temperatures, a characteristic ``one-dimensional'' angular
dependent magnetoresistance is observed, with dips against a broad $cos(\theta )^{\gamma}$ background similar
to (TMTSF)$_{2}$PF$_{6}$.\cite{osada2,iye,caulfield,Kovalev,kartsovnik}  
On the other hand, at high temperature and fields a characteristic ``two-dimensional'' angular
dependent magnetoresistance, with peaks periodic in tan($\theta$)is observed.\cite{ishiguro,house} 
A similar $B-T$ phase diagram is found in the M = Tl and Rb compounds as well.

In this work, we examine the temperature dependence of the magnetoresistance of
$\alpha$-(ET)$_{2}$KHg(SCN)$_{4}$ and $\alpha$-(ET)$_{2}$TlHg(SCN)$_{4}$
as a function of field orientation. Although the compounds are characterized by a complex ground state which is field and
temperature dependent, we find that many aspects of a standard, semi-classical
Boltzmann treatment provide a good account of the data without invoking unconventional transport mechanism.
From the temperature dependence of the magnetoresistance, two features
are evident. The first feature is common to these materials and
to any system exhibiting angular dependent
magnetoresistance oscillations (AMRO). By applying a magnetic field,
a nonmetallic behavior results in the temperature dependence of the
magnetoresistance at low temperatures (this behavior is not indicative of a phase
transition or non-Fermi liquid behavior). Based on the specific electron trajectories along the warped Fermi surface,
there is an effective reduction of the electronic dimensionality. The angular dependence of the electronic dimensionality is reflected in 
the field and temperature dependence of the magnetoresistance. By changing the field orientation,
the temperature dependence of the magnetoresistance oscillates between
that of metallic $(dR/dT > 0)$ at AMRO minima and nonmetallic $(dR/dT < 0)$ at AMRO maxima. 
Because AMRO effects only appear at temperatures below $\approx$ 15 K, we will restrict the temperature range to
25 K and below. The second feature is an angular dependent metallic behavior observed in $\alpha$-(ET)$_{2}$KHg(SCN)$_{4}$ and
$\alpha$-(ET)$_{2}$TlHg(SCN)$_{4}$ inside of DWII.

{\bf II. REVIEW OF SEMICLASSICAL FORMALISM}

For a general anisotropic metal, Fermi liquid theory predicts the phonon scattering rate, 1/$\tau_{p}$, to be proportional to $T^{2}/t _{\bot}$,
where $ t _{\bot}$ is the transfer integral in transverse direction.
The electron-electron scattering rates change with dimensionality. For three dimensions 1/$\tau_{u}$ is expected to be proportional $T^{2}/E_{F}$,
for two dimensions proportional to $(T^{2}/E_{F})ln(E_{F}/T)$
and for one dimension proportional to $T^{2}/t_{b}$, with $E_{F}$ being the Fermi energy.\cite{lev}
Although exact details of the Fermi surface and scattering mechanism (Umklapp, phonon, or defect)in $\alpha$-(ET)$_{2}$KHg(SCN)$_{4}$ is debatable,
we can approximate $\tau$ based on the temperature dependence of the resistance, where $\sigma_{ZZ} \propto constant\times \tau$. At zero fields, the resistance of $\alpha$-(ET)$_{2}$MHg(SCN)$_{4}$ exhibits a metallic
behavior at low temperatures (see Fig. 2). Inside the normal metallic state, the resistance is proportional to the temperature,
$R(T)=A_{1}+A_{2}T$, where $A_{i}$ are constants. Inside of DWI,$R(T)=A_{3}+A_{4}T+A_{5}T^{2}$. 
Note the similarity between R(T) in DWI and in (TMTSF)$_{2}$PF$_{6}$ ($R(T) \propto T^{2}$).\cite{lev}  
By solving the semiclassical Boltzmann transport equation\cite{ziman}, for the specific energy dispersion relation,
the observed AMRO features are
reproduced for both quasi-one and two-dimensional systems.\cite{osada2,iye,blundel,osada3,yagi,danner,moses,schofield,peschansky}
From the calculated conductivity, the temperature dependence can be
predicted at specific field orientations and can be summarized as following.

 {\bf Quasi-one-dimensional} The assumed quasi-one-dimensional energy band dispersion for a pair of warped Fermi sheets
is $\varepsilon_{k}=\hbar \upsilon _{F}(|k_{x}|-k_{F})-(\sum _{m,n}[t^{even}_{mn}cos({\bf R}_{mn} \cdot {\bf k}_{\|})+ t^{odd}_{mn} sin({\bf R}_{mn}\cdot {\bf k}_{\|}$)], 
where $t_{m,n}$ is the transfer integral associated with the oblique lattice vector ${\bf R}_{mn}=(0, mb+nd, nc)$ and describes the warping
topology.\cite{blundel} The conductivity for field rotations in the plane of the Fermi sheets as generated by the semiclassical Boltzmann equation can
be written as\cite{blundel}:
\begin{equation}
\sigma_{ZZ}=N(\varepsilon_{F})\sum_{m,n}^{\infty}\left(\frac{et_{m,n}}{\hbar}\right)^{2}n^{2}c^{2}\frac
{\tau}{1+(G_{m,n}\nu_{F}\tau)^{2}},
\end{equation}
where $N(\varepsilon_{F})$ is the density of states at the Fermi level per
unit volume, $\tau$ is the scattering time, $G_{m,n} = eB((mb+nd)cos \theta -ncsin \theta )/\hbar$, $B$ is the magnetic
field, $e$ is the electron charge, and $\nu_{F}$ is the Fermi velocity.
The $G_{m,n}$ term contains the field dependence of $\sigma_{ZZ}$.

Although no semiclassical prediction for the temperature dependence of the magnetoresistance in a quasi-one-dimensional case has been made, 
it is easily derived. When the field orientation is aligned at AMRO minima,
$G_{m,n}$, vanishes and the conductivity is simply $\sigma_{ZZ}=constant\times \tau = constant'\times \sigma_{ZZ}(B=0)$,
resulting in a temperature dependence of
the resistivity being metallic(dR/dT $>$ 0). On the other hand, at AMRO maxima, $G_{m,n}$ dominates and the
conductivity can be written as $\sigma_{ZZ}=1/(G_{m,n}\nu_{F})^{2}\tau$. The resulting temperature dependence of the resistivity at AMRO
maximum is proportional to $\tau$ and is nonmetallic(dR/dT $<$ 0). From the zero field data we know that in the DWI state that 
$1/\tau$ is proportional to R(T), we can easily predict the temperature dependence of the resistivity;
at AMRO maxima $R(T) \propto 1/(a_{1}T+a_{2}T^{2})$, 
and at AMRO minima $R(T) \propto a_{3}T+a_{4}T^{2}$, where $a_{i}$ are constants. Furthermore, $G_{m,n}$ has a linear field dependence.
As the magnetic field strength increases, the resistance at AMRO maxima will increase, whereas the magnetoresistance at the AMRO minima will be almost field independent.
Thus by increasing the field strength the metallic/non-metallic behavior becomes more pronounced. 

{\bf Quasi-two-dimensional} The assumed quasi-two-dimensional energy band dispersion for a corrugated Fermi cylinder is
$\varepsilon_{k}$=$\hbar$$^{2}(k_{x}^{2}+k_{y}^{2})/2m^{*}-2t_{z} cos(ck_{z})$. 
Here $k_{x}$ and $k_{y}$ are the components of the wave vector $\bf k$ in the conducting-plane,
c is the inter-plane distance, $k_{z}$ is the wave vector component
normal to the planes, and m$^{*}$ is the effective cyclotron mass in the conducting
plane. The interlayer transfer integral $t_{z}$ is small compared to the Fermi
energy such that $\varepsilon_{F}/t_{z} \gg 1 $, generating a slightly warped cylindrical
Fermi surface. By solving the Boltzmann 
equation, the conductivity (or resistivity) along the least conductive axis can now be calculated.
When the field is sufficiently high or the temperature is sufficiently low, $\omega_{o} \tau cos( \theta _{max}) >>$ 1 , where $\omega _{o}$ is
the cyclotron frequency, $eB/m^{*}$, 
then at AMRO maximum the normalized resistivity can be written as\cite{moses}:

\begin{equation}
\frac{\rho_{zz}(\theta_{AMRO max})}{\rho_{zz}(B=0)}=\frac{\gamma(\omega\tau)^{2}sin(2\theta_{AMRO max})}{\pi}.
\end{equation}
The resulting temperature dependence of the resistivity at AMRO
maximum is proportional to $\tau$ and is nonmetallic (dR/dT $<$ 0).

At the AMRO minimum, the normalized resistivity can be written as\cite{moses}:
\begin{equation}
\frac{\rho_{zz}(\theta_{AMRO min})}{\rho_{zz}(B=0)}=\frac{\pi^{2}}{2}\left(n+\frac{1}{4}\right),
\end{equation}
where $n$ is an integer. The temperature dependence of
the resistivity at AMRO minima is proportional to 1/${\tau}$ and metallic(dR/dT $>$ 0). 
The temperature dependence has approximately the same behavior as in the 
quasi-one-dimensional case, however in this case $1/\tau$ is proportional to $T$ instead of $T+T^{2}$.

The above discussion is also understandable in a rather intuitive way.
By changing the field orientation, the electron trajectories along the warped Fermi surface also change.
For specific field orientations the velocity is more effectively averaged to zero and transport is reduced. 
For the Fermi cylinder, the velocity vector along the cylinder axis goes to zero when all closed orbits share the same area.
For the Fermi sheets, the velocity vector is reduced when the electrons are not traveling along the axis of corrugation.
Thus, the effective dimensionality oscillates with field orientation. At AMRO maximum the system has reduced
dimensionality and at AMRO minimum the dimensionality is restored.

{\bf III. EXPERIMENT}

Single crystals of $\alpha$-(ET)$_{2}$MHg(SCN)$_{4}$, where M = K or Tl,
were grown using conventional electrocrystallization techniques.\cite{ishiguro} Systematic
measurements of the temperature dependence of the magnetoresistance from
30 K to 1.5 K at various field orientations and strengths were performed
in the 33 tesla resistive magnet at NHMFL, Tallahassee. The magnetoresistance
was measured using standard four terminal AC techniques with 12 $\mu$m gold wires attached
via graphite paste. The current was applied along the least conducting ${\bf b}$$^{\star}$ axis. Typical contact resistance was below 10 $\Omega$. 
Samples were rotated in magnetic field with $\theta$ being defined as the angle between the
${\bf b}$$^{\star}$ axis and field. The field projection angle($\phi$) with respect to the ${\bf a}-{\bf c}$ plane was determined via polarized
infrared reflectance. The magnetic field was oriented $\approx 8^\circ$ from 
the ${\bf c}$ axis in the M = K sample and  $\approx 30^\circ$ from ${\bf c}$ axis in the M = Tl sample.

{\bf IV. RESULTS}

{\bf A. Temperature dependence in DWI}

AT 14 T the electronic state is well inside DWI. Typical AMRO
for this regime is shown in Fig. 3(a). The oscillations are characterized by minimum periodic in
tan($\theta$) and a maximum when the field is near the {\bf b$^{\star}$} axis.\cite{osada2,iye,caulfield,Kovalev,kartsovnik}
The background magnetoresistance can be fit
to $cos(\theta )^{\gamma}$, where $\gamma$ varies from 0.6 to 1.6 based on sample quality and field strength.
The AMRO is very similar to that observed in (TMTSF)$_{2}$PF$_{6}$ where the background magnetoresistance 
can be fit to $cos(\theta )^{\gamma}$, where $\gamma$$\approx$ 0.5,\cite{kang} however the dips in (TMTSF)$_{2}$PF$_{6}$
are not as deep as in $\alpha$-(ET)$_{2}$MHg(SCN)$_{4}$.  The temperature dependence of the
magnetoresistance for different field orientations for $\alpha$-(ET)$_{2}$KHg(SCN)$_{4}$ are shown in Fig. 3(b). 
When the field direction corresponds to an AMRO minimum, the sample resistance is metallic
$(dR/dT > 0)$ (see Fig. 3(b) curve D). Deviations from this orientation results in nonmetallic $(dR/dT < 0)$ behavior in 
the temperature dependence. This nonmetallic behavior becomes more pronounced as the field orientation
approaches AMRO maxima (curves C, B, and then A). In some samples the
nonmetallic behavior, occurring at AMRO maxima saturates at very low temperatures. This saturation of the magnetoresistance arises
due to impurities or lattice defect scattering and indicates the limit at which the dimensionality can be reduced. 

Figure 4 is a plot of the temperature dependence of the magnetoresistance
in $\alpha$-(ET)$_{2}$KHg(SCN)$_{4}$ at AMRO maximum and
minimum (as indicated in Fig. 3) in a wide temperature range. The onset of DWI is very clear. 
The data is represented by open circles and the fits by solid lines. The fits are obtained from the semiclassical
Boltzmann treatment described in the previous section and using the zero field temperature dependence to determine $1/\tau$ (AMRO minimum fit to $A_{1}+A_{2}T+A_{3}T^{2}$ and AMRO maximum fit to 
$A_{4}/(A_{5}+A_{6}T+A_{7}T^{2})$, where $A_{i}$ are constants).
At field orientations corresponding to AMRO
maximum, the sample resistance is proportional to a $1/(a_{1}T+a_{2}T^{2})$  
dependence. At field orientations corresponding to AMRO minimum, the
temperature dependence of the resistance is the same as that realized at
zero fields and is proportional to the temperature, $a_{3}T+a_{2}T^{4}$, where $a_{i}$ are constants. 

Although the angular dependent magnetoresistance oscillations can be
reproduced via semiclassical descriptions, there are still some complications to be
addressed. It has been shown that
the magnetoresistance violates Kohler's rule in DWI, thus indicating that semiclassical methods
may not be valid.\cite{spuds2} It is likely that there are additional complications 
such as magnetic breakdown, electronic subphases, or mixed states inside of DWI. 

Another approach to explain AMRO in the DWI of $\alpha$-(ET)$_{2}$KHg(SCN)$_{4}$
considers incoherent transport.\cite{yoshioka} Albeit this approach is not strictly semiclassical, it is
included for completeness. In this model the reconstructed Fermi
surface of DWI is not characterized by a quasi-one-dimensional
but a quasi-two-dimensional topology. In fact, one of the proposed models for the reconstructed Fermi surface 
inside of DWI has the Fermi surface characterized by cylinders
instead of sheets. The alternative
model for the angular dependent magnetoresistance can be summarized in the following way. When
a magnetic field is applied to a density wave structure, a resulting periodic potential is
created in the inter-plane direction. As the field direction is
rotated, this periodicity changes. If the periodicity becomes
commensurate with the interlayer spacing, the electron states are
extended. This corresponds to AMRO minima. At the minima, the electronic system 
is metallic and the sample resistance should decrease
with decreasing temperatures. When the inter-plane
potential is not commensurate to the interlayer spacing the
magnetoresistance increases. When the potential is exactly out of phase
with the interlayer spacing the system now corresponds to AMRO
maxima. In this orientation, the electron wave function begins to
shrink and localize with increasing field or decreasing
temperature, resulting in incoherent electron motion. In this
field orientation, the sample resistance should increase with
decreasing temperature. Unfortunately, no specific predictions for the
functional form of the temperature dependence have been made yet for
this model.

{\bf B. Temperature dependence in the normal metal state}

At 30 and 33 tesla, $\alpha$-(ET)$_{2}$MHg(SCN)$_{4}$ is in the normal
state except for angles $\theta$ $\geq 80^\circ$ and for temperatures
below $\approx$ 3 K (inside DWII). Figure
5 (left side) displays typical quasi-two-dimensional AMRO as observed in the
normal metal state and DWII. The oscillations are characterized by maxima periodic in
tan($\theta$) with the first minimum occurring near $\theta$ = 0 $\pm$
$6^\circ$.\cite{yamaji} We have measured the temperature dependence of R(T) at many field orientations. 
The solid arrows indicate field orientations (E, F, G, and H as shown in Fig. 5 right side) where the temperature dependence
of the magnetoresistance was measured. Unfortunately, 
Shubnikov-de Haas (SdH) oscillations dominate as the sample is rotated in high magnetic fields.\cite{shoenberg} The SdH oscillations
are superimposed on top of the angular dependent magnetoresistance oscillations. If the field orientation corresponds
to either SdH maximum or minimum, an additional term will appear in the temperature
dependence of the magnetoresistance. Therefore field orientations
corresponding to SdH minimum or maximum will be omitted in the
following discussion by avoided angles near $\theta$=0$^\circ$. 

Inside the normal state we find that when the field is aligned along AMRO minimum, the temperature dependence of the magnetoresistance is
metallic${(dR/dT > 0)}$ (see right side Fig. 5 curve H). Deviations from field directions corresponding
to precise minimum position results in a change of the temperature dependence ${(dR/dT < 0)}$. This nonmetallic behavior 
is more pronounced near AMRO maximum (see right side Fig. 5 curves G, F, and then E). Although the AMRO has changed its form, the
temperature dependence is very similar to Fig. 3 and 4. 

The temperature dependence of the magnetoresistance at the AMRO maximum
and AMRO minimum is displayed in Fig. 6 (the field orientations are indicated in Fig. 5). The solid lines are fits (AMRO minimum fit to $A_{1}T+A_{2}$ and AMRO maximum fit to 
$A_{3}/(A_{4}T$+$A_{5}T^{2})$, where $A_{i}$ are constants). Like inside DWI the temperature dependence is described by semiclassical Boltzmann treatment.
Deviations to the fit begin to occur at lowest temperatures as DWII and a new metallic behavior sets in (see Fig. 5 curve E).

At first glance one might believe that the nonmetallic like behavior is resulting from a gap opening. Yet, the behavior of the temperature dependence
can be explained in terms of semiclassical transport. At AMRO maximum, the system is more
effectively two dimensional with increasing field strength or decreasing temperature. All quasi-two-dimensional systems exhibiting 
AMRO will experience nontrivial temperature dependence.

{\bf C. Temperature dependence in DWII}

When the magnetic field orientation is close to AMRO
maxima a new behavior is observed at high fields and low
temperatures in the temperature dependence of the magnetoresistance. There is a transition from a non-metallic to a metallic behavior
at a transition temperature $T_{M}$ (see right side Fig. 5 curve E). This reentrant 
metallic behavior is a signature of the DWII regime.  
In one sample of
$\alpha$-(ET)$_{2}$TlHg(SCN)$_{4}$, $T_{M}$ was $\approx$ 4 K and could be observed at many angles, especially near AMRO maxima.
Figure 6 shows the evolution of the slope and $T_{M}$ with
respect to $\theta$ at 30 and 33 T. Once inside DWII, both the magnetoresistance and the
magnetization drop abruptly, however, the mechanism responsible for the behavior is questionable.\cite{kartsovnik2,christ}
From AMRO and Shubnikov-de Haas  measurements, the Fermi cylinders
of the normal state are clearly observed within DWII,\cite{house} suggesting that the Fermi surface is very similar to that of the normal state. 
The replacement of the nonmetallic behavior by the metallic one indicates the existence of a new unidentified property of DWII,  and may be related to
the appearance of large eddy currents observed in previous experiments
within the DWII regime.\cite{honold,harrison,yaguchi}

If we assume the electronic conduction in DWII is due to the normal state plus that of another highly conductive transport channel,
we are able to fit the measured data. Two contributions,
one due to the original Fermi surface (an angular dependent term
$\propto$ 1/T) and the other due to the new metallic term (an angular independent term $\propto$ T), are
added in parallel. The fits are shown in Fig 6(b). This reproduces the
observed behavior and describes the dependence of $T_{M}$ on
field orientation and strength. The slope of the temperature
dependence of the magnetoresistance will also be greater at AMRO maximum. This is because there is a significant difference 
in the magnetoresistance between the AMRO maximum and minimum (see left side Fig. 5). 
The origin of this new metallic term is not clear. A possible explanation is a transfer of carriers from the 
original quasi-two dimensional Fermi surfaces to another conduction channel. By removing carriers from the cylinders,
yet retaining them on some other Fermi surface, there is an increase of dimensionality and conductivity. 
Could it indicate the existence of highly conductive edge states or the emergence of a new Fermi surface? 
With the onset of DWII, the Fermi surface reconstructs and the effects of the
original Fermi surface are being suppressed (including the nonmetallic behavior). At high fields or specific field orientations,
the nesting condition may improve such that more carriers are transferred from the original Fermi
surface. Further investigations of this unusual phase are required.

{\bf V. SUMMARY}

We have investigated the effects of field strength and orientation
on the temperature dependence of the magnetoresistance in samples of
$\alpha$-(ET)$_{2}$MHg(SCN)$_{4}$ (M = K or Tl). At angular dependent
magnetoresistance oscillation (AMRO) minima the temperature dependence is metallic
and at AMRO maxima the magnetoresistance shows nonmetallic temperature dependence.
When the magnetic field is oriented along 
AMRO maxima, the effective electronic dimensionality decreases with
decreasing temperature and increasing field.  We show that non-Fermi liquid transport
is not necessary to generate an angular dependent temperature dependence. 
This behavior can be explained in terms of semiclassical electron
transport and the Boltzmann equation. Although the scattering rates for every system will be different based on details of the Fermi surface and what not,
the metallic/nonmetallic temperature dependence of the magnetoresistance should be common to any low dimensional 
system with AMRO.

The temperature dependence of the magnetoresistance at different fields (for fixed angle) 
reported by Ref.\cite{kartsovnik2} and Ref.\cite{sasaki} on $\alpha$-(ET)$_{2}$KHg(SCN)$_{4}$ can also be explained in terms of semiclassical theory. 
In the first reference the field is oriented at an AMRO minimum in DWI and an AMRO maximum in the normal state (see Fig. 1 Ref. \cite{kartsovnik2}).
The temperature dependence of the magnetoresistance is non-metallic in the normal state, and metallic in DWI and DWII. 
In the second reference, Ref. \cite{sasaki} the field is oriented near $\theta$=0$^\circ$ or at an AMRO maximum in DWI and at an AMRO minimum in the normal state.
The temperature dependence of the magnetoresistance is metallic in DWII,
however there is a nonmetallic behavior in DWI {\em and} the normal state. Albeit the effect is less pronounced in later measurements by the 
same author, the nonmetallic behavior in the normal state is reproduced(at the same field orientation).
It is not known if this is an effect particular to $\theta\approx$ 0$^{\circ}$ or not.
The field orientation may not have been along the precise AMRO minimum.
If the sample is slightly misalign or the first AMRO minimum is not at $\theta$= 0$^\circ$ (there is a range of $\pm$
$6^\circ$ based on the $\phi$ field orientation in the conducting plane), the sudden rise before DWII state
would correspond to the non-metallic behavior observed when away from precise AMRO minima. 

A quasi-one-dimensional organic charge
transfer salt (TMTSF)$_{2}$PF$_{6}$ exhibits a very similar temperature dependent magnetoresistance
as that in the low temperature regime of $\alpha$-(ET)$_{2}$MHg(SCN)$_{4}$.\cite{chashech}
The temperature dependence of (TMTSF)$_{2}$PF$_{6}$ has been modeled extensively
in terms of non-Fermi liquid transport. It would be interesting to test semiclassical models on this temperature dependence
to see if non-Fermi liquid behavior is required.

To first order, the transport in all regimes of
$\alpha$-(ET)$_{2}$MHg(SCN)$_{4}$ can be
understood in terms of semiclassical orbits(using both quasi-one and quasi-two-dimensional models). 
This semiclassical approach leads, naturally, to a nonmetallic behavior at all 
maxima in angular dependent magnetoresistance oscillations due to a reduction of the
effective electronic dimensionality with increasing magnetic field or
decreasing temperature. Furthermore, a new metallic property of DWII is reported. 
The mechanism responsible for this metallic behavior is
unknown and the nature of the DWII phase will be the subject of future investigations.

{\bf ACKNOWLEDGMENTS}
We would like to thank L. Balicas, S. Y. Han, and B. H. Ward for useful discussions. This work was supported by the National Science Foundation
under Grant No. NSF-DMR-99-71474.

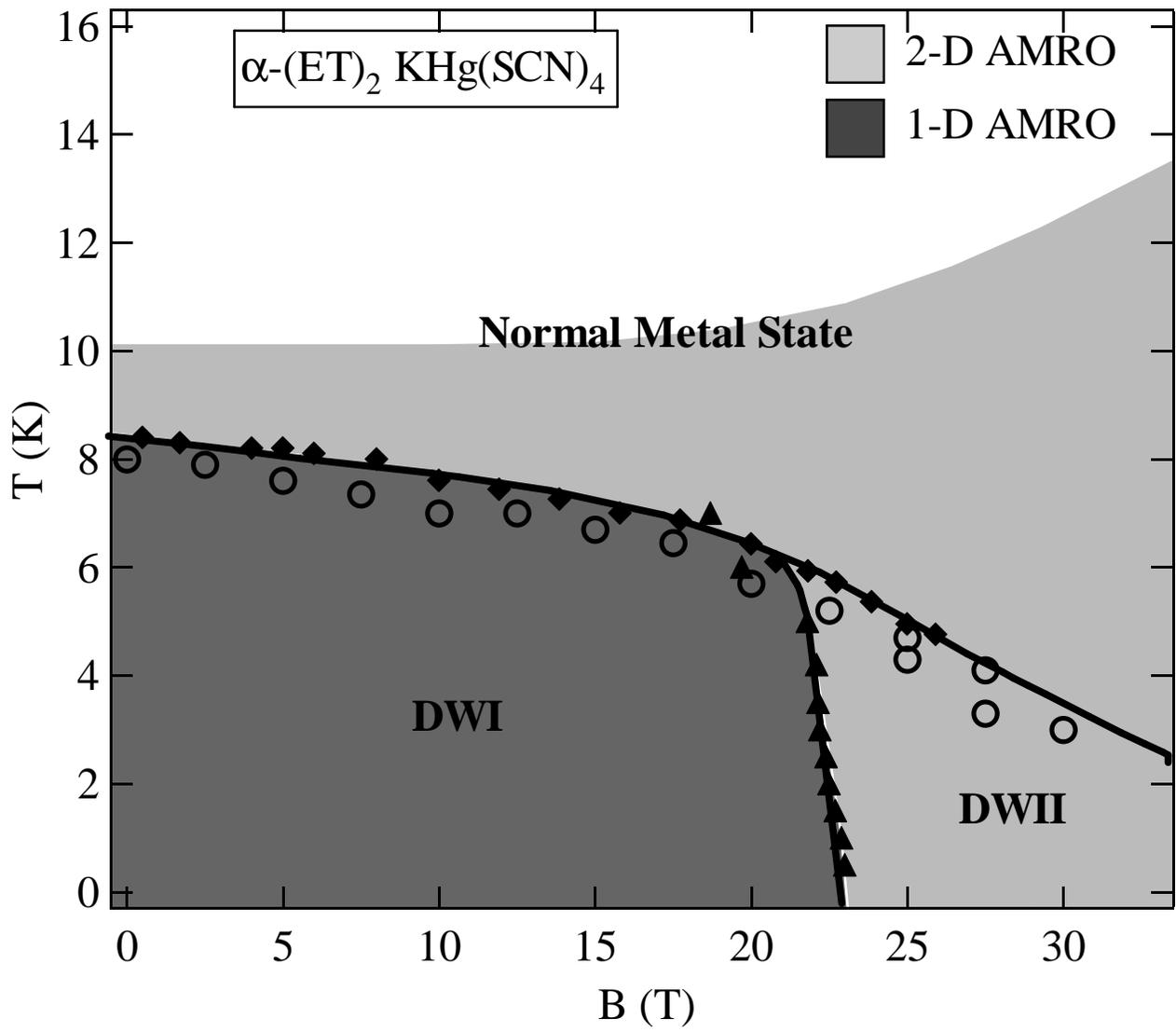
\begin{figure}[htbp]
\caption{Schematic phase diagram and AMRO behavior in the M = K salt. The diamonds [12] and circles [11]
indicate the transition into the low temperature ground 
state from transport measurements. The triangles signify the kink field, $B_{K}$ and separate DWI from DWII. 
The light gray, and dark gray color
corresponds respectively to regions of quasi-two-dimensional or quasi-one-dimensional. For the
hysteretic regions near $B_{K}$ the AMRO behavior is mixed.
A similar diagram is observed for the M = Rb and Tl compounds .} 
\end{figure}

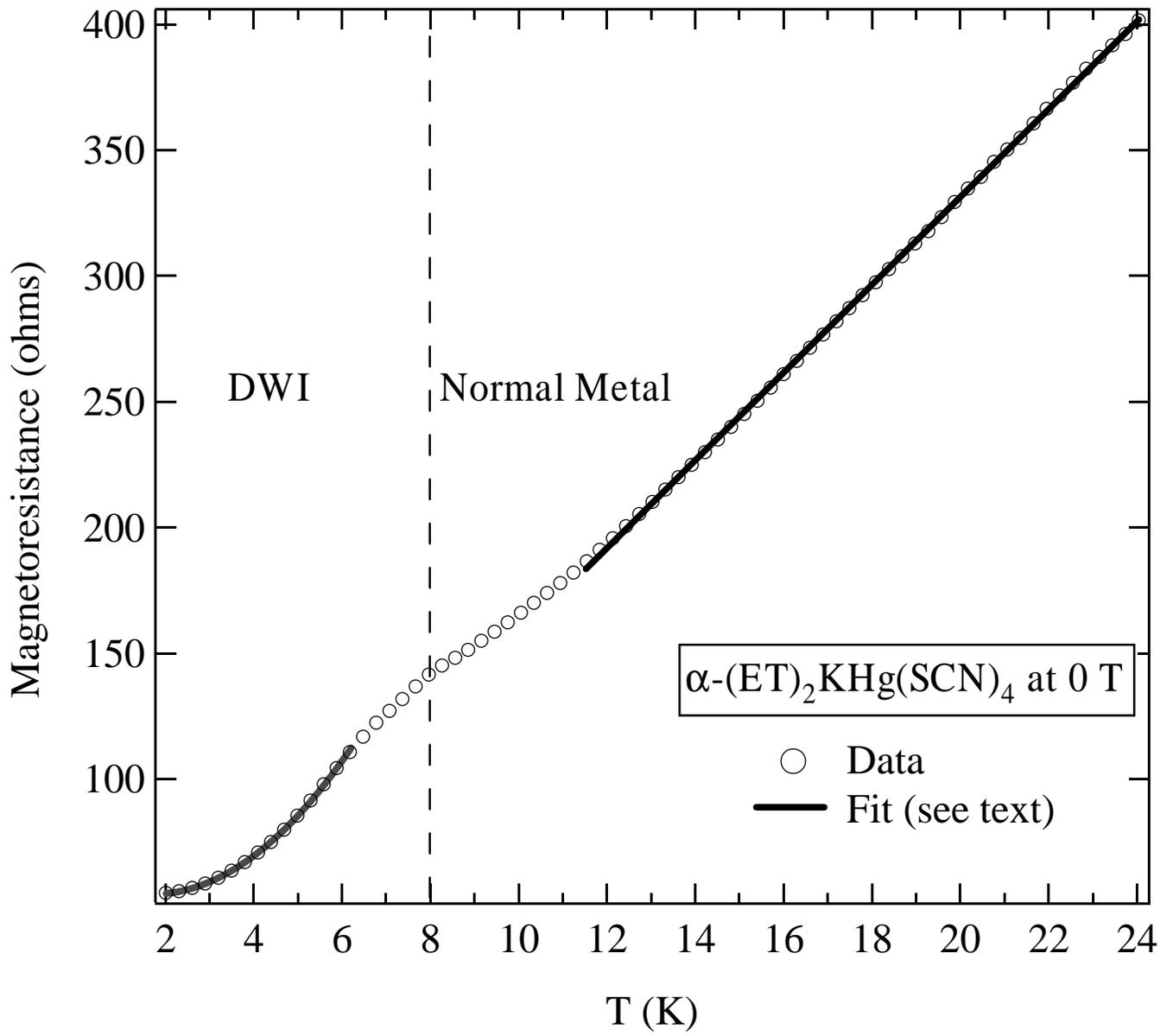
\begin{figure}[htbp]
\caption{Temperature dependence of the resistance in $\alpha$-(ET)$_{2}$KHg(SCN)$_{4}$ 
at zero magnetic field. Inside the normal metal state, $R(T) \propto T$.
 Inside DWI, $R(T) \propto \alpha T+ \beta T^{2}$.} 
\end{figure}

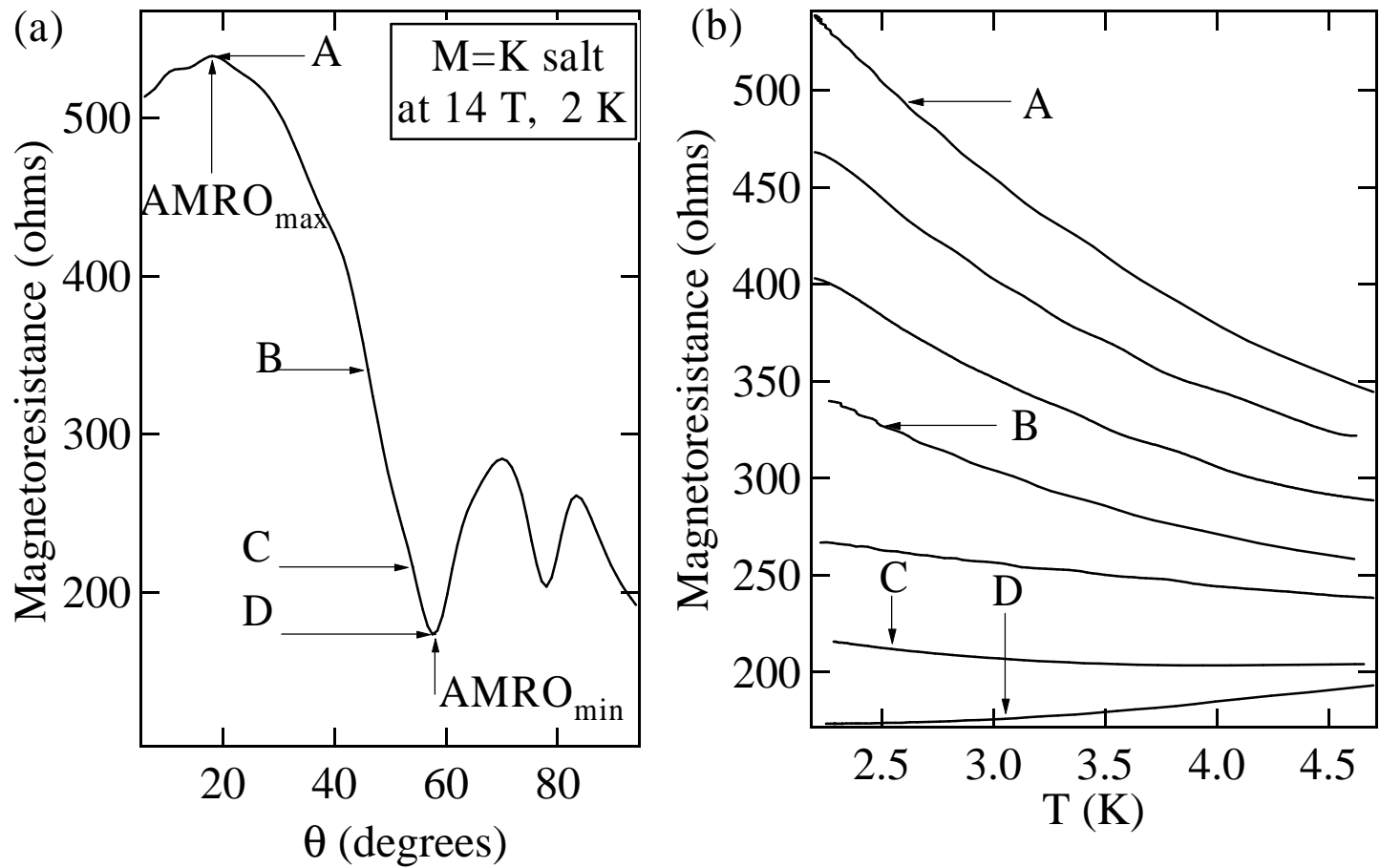
\begin{figure}[htbp]
\caption{(a) AMRO in $\alpha$-(ET)$_{2}$KHg(SCN)$_{4}$ at 14 tesla and 2 K
in DWI. (b) The temperature dependence for
the field orientations marked by the arrows in (a). A metallic temperature dependence is observed when the field
orientation corresponds to AMRO minima (curve D). On the other hand at AMRO maxima 
the behavior is nonmetallic (as seen in C, B, and A). }
\end{figure}

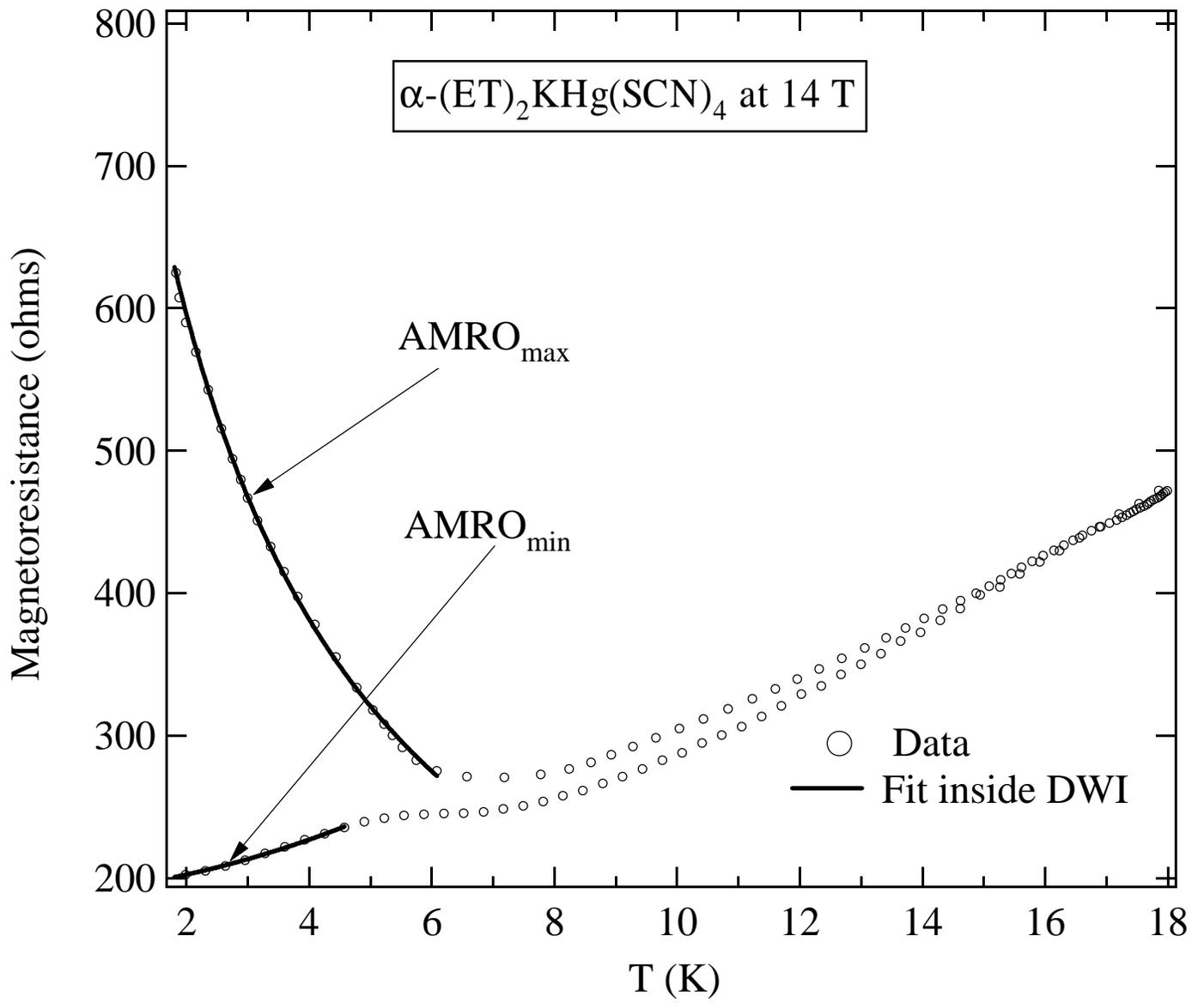
\begin{figure}[htbp]
\caption{Temperature dependence of the magnetoresistance in
DWI at 14 tesla. At AMRO maxima the temperature dependence can be
approximated to be proportional to $\tau$. At AMRO minima the temperature
dependence is approximated to be proportional to 1/$\tau$ and T. R(T) at the AMRO minimum is fit
to $A_{1}+A_{2}T+A_{3}T^{2}$ and at AMRO maximum it is fit to 
$A_{4}/(A_{5}+A_{6}T+A_{7}T^{2})$, where $A_{1}$, $A_{2}$, $A_{3}$, $A_{4}$, $A_{5}$, $A_{6}$, and $A_{7}$ are constants. } 
\end{figure}

\begin{figure}[htbp]
\caption{AMRO in $\alpha$-(ET)$_{2}$KHg(SCN)$_{4}$ at 30
tesla and 0.5 K corresponding to the normal state and DWII (Left side). 
Temperature dependence of
$\alpha$-(ET)$_{2}$KHg(SCN)$_{4}$ magnetoresistance
at various field positions as marked by arrows on the left side (Right side). A metallic
temperature dependence is observed at AMRO minima (curve H) and a nonmetallic
dependence at all other positions (curve G, F, and E). A reentrant metallic behavior is evident 
for positions corresponding to AMRO maxima below 3 K (inside DWII) (curve A).} 
\end{figure}

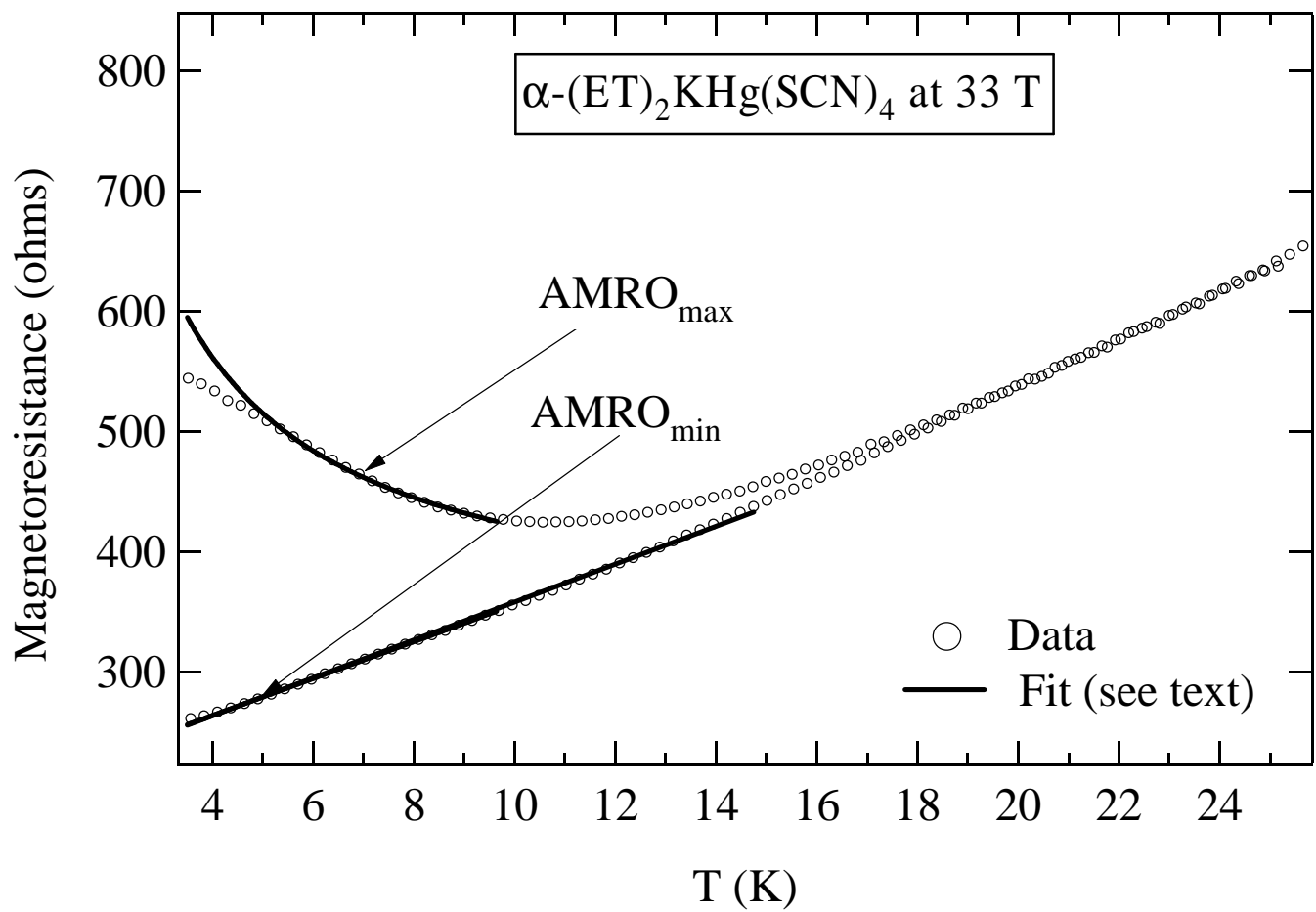
\begin{figure}[htbp]
\caption{Temperature dependence of the magnetoresistance at 33 tesla.
Anisotropic temperature dependence develops along with the AMRO. At AMRO maxima the temperature dependence is inversely proportional to T.
The fit in the figure is $R(T)$=${A}_{1}$/T + $A_{2}$, where ${A}_{1}$ and $A_{2}$ are constants.
At lowest temperatures the resistance starts to saturate and the fit fails. At AMRO minima the
temperature dependence is proportional to T. The fit in the figure is $R(T)$=$A_{3}$T
+ $A_{4}$, where $A_{3}$ and $A_{4}$ are constants.} 
\end{figure}

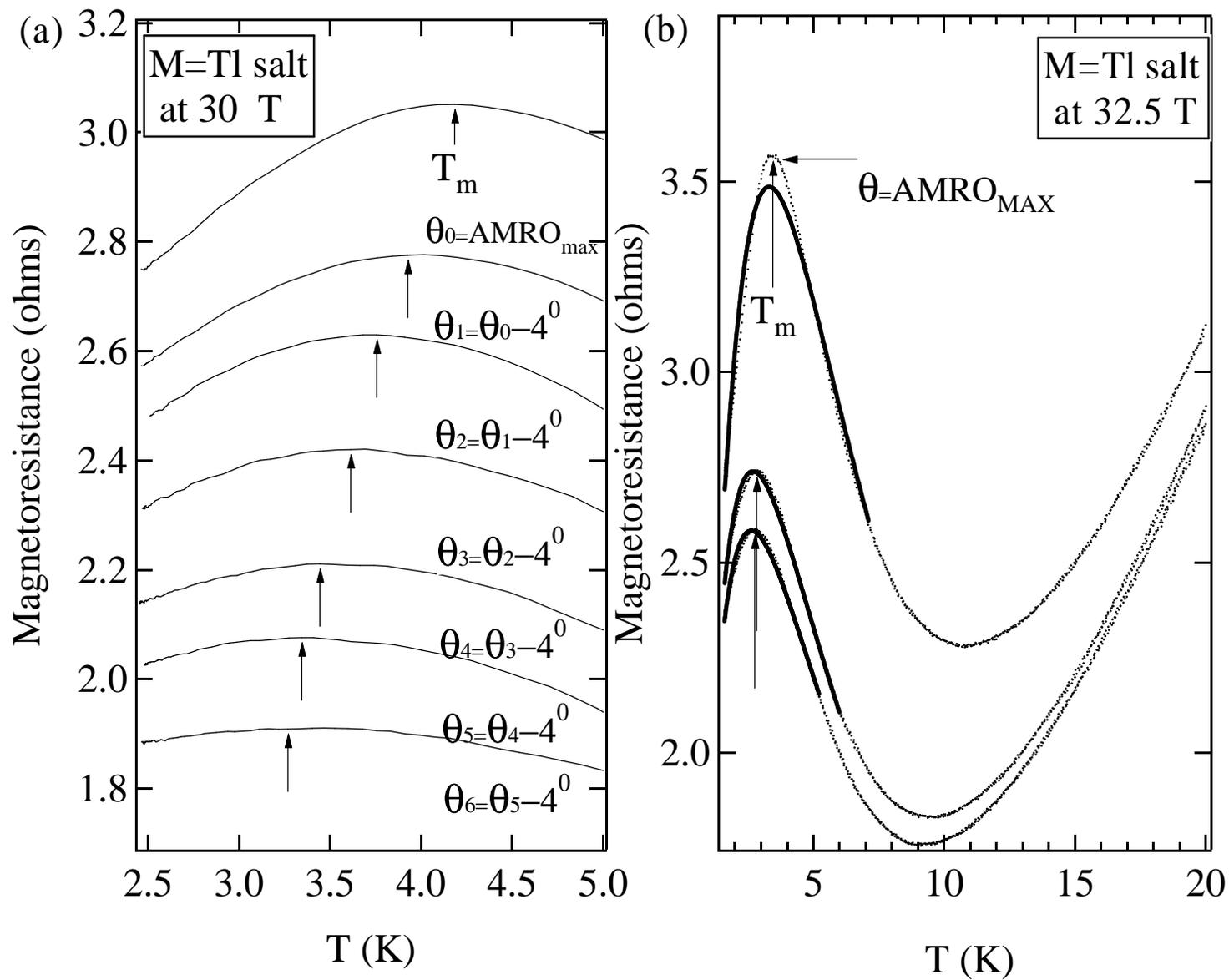
\begin{figure}[htbp]
\caption{Temperature dependence of $\alpha$-(ET)$_{2}$TlHg(SCN)$_{4}$ in
DWII near field orientations corresponding to AMRO maxima,
reentrant metallic behavior is observed. Arrows mark the transition temperature
T$_{M}$ into this new metallic behavior. Both the slope of the temperature dependence and the value of T$_{M}$ change with
respect to field strength and orientation. a) 30 tesla data and b) 32.5
tesla data reveals the behavior clearly. The dashed lines are fits by assuming the conductivity 
is the sum of the $R(T,$$\theta$) of the normal state plus a new angular independent metallic term (see text).} 
\end{figure}

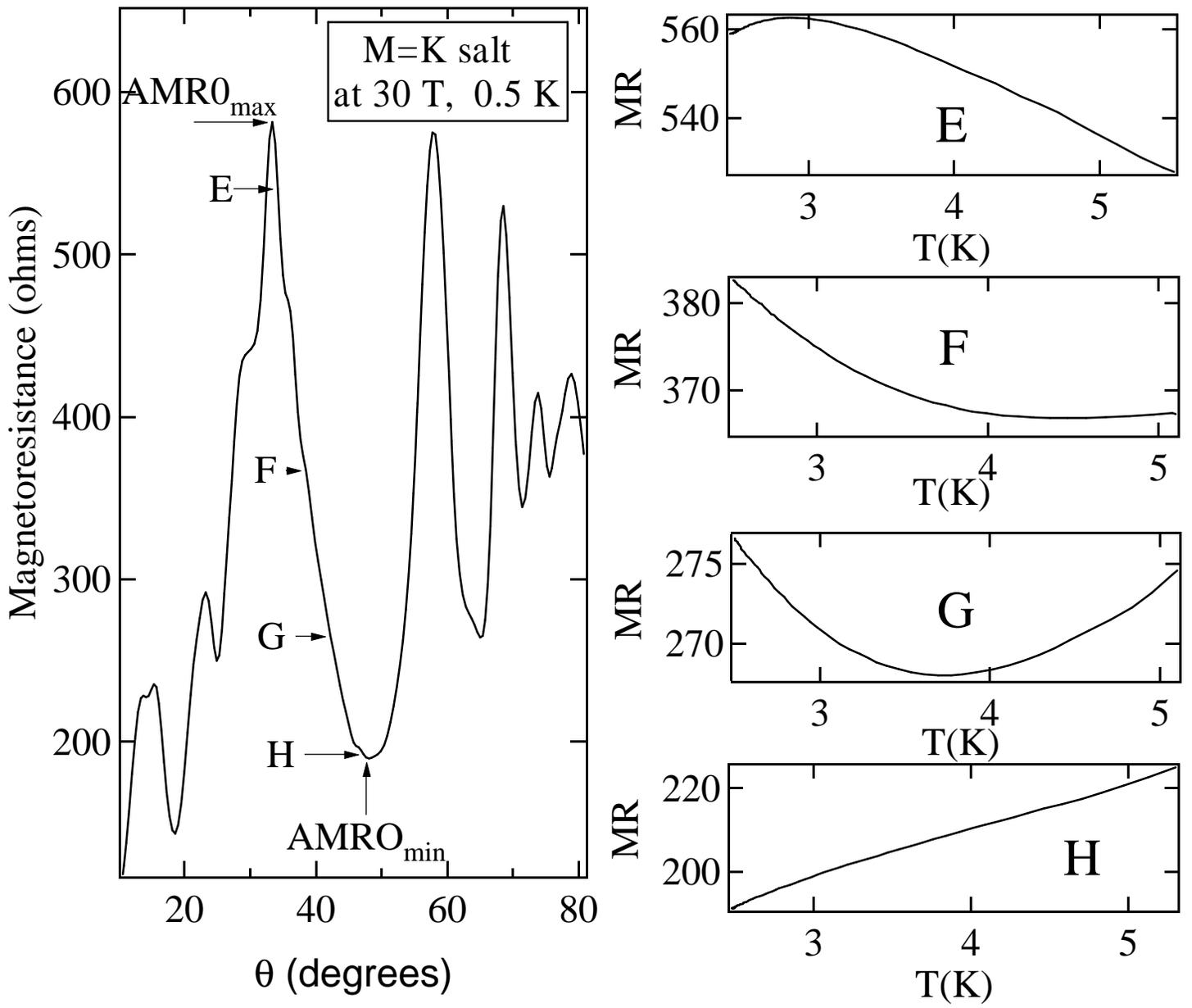
\end{document}